# Smartphones in Mental Health:

## Detecting Depressive and Manic Episodes

Venet Osmani[1], Agnes Gruenerbl[2], Gernot Bahle[2], Christian Haring[3], Paul Lukowicz[2], Oscar Mayora[1]

[1]CREATE-NET, Trento, Italy

[2]DFKI, Kaiserslautern, Germany

[3]TILAK, Hall in Tirol, Austria

As we study human nature, we're realizing that successful disease treatment outcomes depend on incorporating individuals' phenotype into the treatment model. This realization has given rise to personalized medicine, which aims to integrate data on the genetic makeup of individuals as well as environmental and lifestyle factors. Personalized medicine will greatly benefit from genomics [1], but to build a complete picture of individuals' phenotype, we also need to better understand environmental and behavioral factors.

Specifically, behavioral data will have a significant impact on our understanding of mental disorders. Because symptoms of most mental disorders are manifested as changes in an individual's behavior, analyzing such changes could lead to a better understanding of these types of diseases and possible treatments. Given that we don't yet have strong biomarkers for diagnosing most mental disorders, behavior evaluation based on subjective information remain the primary means of diagnosis.

Recently, behavior understanding has been given a boost from rapid increases in the capabilities and miniaturization of sensing devices. Specifically, the evolution of smartphones from communication-centric devices toward devices with multimodal sensing capabilities promises to provide unprecedented insights into human behavior (see the "Smartphone Sensing" sidebar for more information). Motivated by the advantages of using smartphones for behavior monitoring in general and monitoring of symptoms of neurological disorders in particular, my collaborators and I from the MONARCA project (http://www.monarca-project.eu) carried out an observational study with patients diagnosed with bipolar disorder. Bipolar disorder is a common and severe form of mental illness characterized by repeated relapses of mania and depression. Patients suffering from the disorder thus can experience—often in rapid succession—periods of manic, normal, and depressive states.

The current standard for diagnosing bipolar disorder uses subjective clinical rating scales based on self-reports. Scales such as the Hamilton Depression Rating Scale (HAMD) were developed in the early 1960s, and there are also more recent variations, such as the Bipolar Spectrum Diagnostic Scale (BSDS). Although the efficacy of these scales has been proven in diagnosing bipolar disorder, they have their drawbacks because of the potential for subjectivity in the diagnosis. To address this issue, the aim of our study was to investigate whether data from smartphone sensors could be used to recognize bipolar disorder episodes and to detect behavior changes that could signal the onset of an episode using objective, sensor data.

## Study Overview

The study [2] was set up in a psychiatric hospital in Hall in Tirol, Austria and approved by the ethics board of the Innsbruck University Hospital. It lasted from November 2012 to August 2013 and involved 12 patients. Each patient was continuously monitored during his or her daily life for 12 weeks on average, resulting in over 1,000 days of smartphone sensor data. There were no constraints of any kind placed on the patients with respect to holding the phone in a specific manner or at a specific place on the body or otherwise. Each patient underwent a mental state examination at the beginning and end of the study and every three weeks in between; more frequent examinations would have resulted in a learning effect that could have biased the outcome. Mental state examinations resulted in patients' state score that was normalized between −3 (episode of severe depression) and +3 (episode of severe mania), having moderate and mild conditions represented at scales ±2 and ±1, respectively. Figure 1 shows the evolution of the patients' mental states during the monitoring period. (Note that two patients, p0202 and p0602, withdrew early, and patients p0402 and p0202 didn't experience a change of state during the monitoring period and thus weren't considered in the analysis.

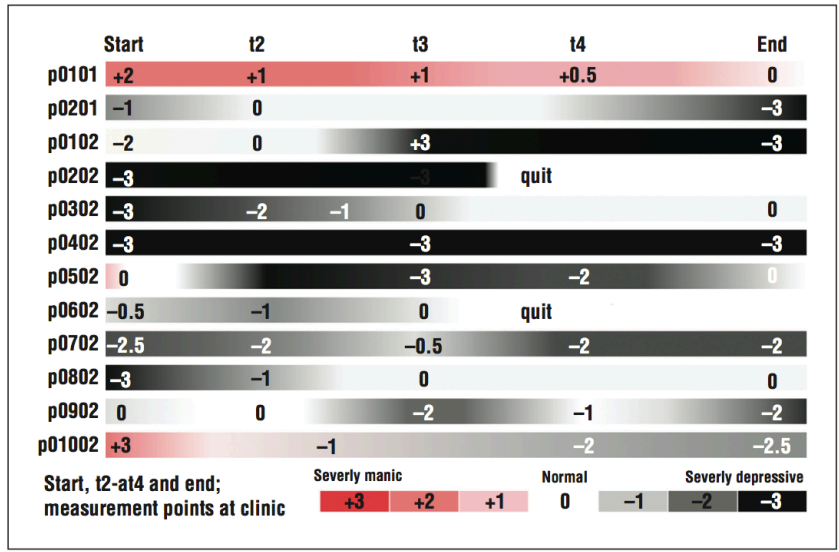

*Figure 1. Mental state of the patients during various stages of the monitoring period* [2]

# Results

Considering that depressive and manic episodes are manifested through psychomotor retardation and agitation respectively, as the first step, we investigated the correlation between physical activity (measured through the smartphone's accelerometer) and patient states [3]. Afterward, we investigated the potential of smartphone sensor data to recognise a bipolar episode and detect changes in a patient's state that could signal the onset of an episode.

## Initial Analysis of Physical Activity Data

A daily physical activity score was calculated for each patient, excluding days when the patients went to the clinic for their mental state examination. We excluded these days so as not to bias the results, because on these days, physical activity would have been present independent of the patient's condition.

Surprisingly, the Pearson product-moment correlation coefficient between daily physical activity scores and patients' state showed weak correlation ($r = 0.3638$, $p < 0.05$). One of the reasons for the weak correlation was that a daily activity score didn't capture individual differences between patients, both in terms of circadian rhythms as well as other patterns of daily activities.

Upon receiving advice from the psychiatrists, we divided the day into four intervals (morning, afternoon, evening, and night), and we calculated an activity score for each interval. This resulted in much stronger correlation between daily intervals' activity scores and patients' mental state examination scores ($r = 0.6248$, $p < 0.05$). This initial investigation was then followed by further analysis of data from other sensing modalities with the aim of recognizing bipolar episodes and detecting changes that might signal the onset of an episode.

## Recognizing Bipolar Disorder Episodes

Although the correlation between physical activity and bipolar episodes has been established in the medical literature, other aspects of the behaviour of bipolar disorder patients haven't previously been measured. These aspects include location and mobility patterns, voice analysis during phone calls, and analysis of phone call patterns. Our subsequent study investigated whether accelerometer and GPS data could be used to recognize the patient state and detect the onset of an episode.

**Recognizing the patient state.** Considering that ground truth data (a psychiatric mental state examination) was at three-week intervals, we chose a period of seven days before and two days after the mental state examination for the sensor data. This was based on the assumptions elicited from discussions with the psychiatrists that state changes are gradual and the probability of a major change within a few days is low.

Extracted features from accelerometer signals and GPS traces were used in single modality classification of the patient state and were also fused together [2], using mental state examination as the ground truth. A within-patient Naïve Bayes classification achieved 81% mean accuracy in recognizing the patient state (k-nearest neighbour, j48 search tree, and a conjunctive rule learner yielded similar performance results). The classifier's precision was 81%, and recall was 82%.

As Table 1 shows, patients' location data is a good predictor of their state and, in our case, it was even better than physical activity (obtained from the accelerometer data) for predicting the state [2].

**Table 1. Classifier accuracy performance in recognizing the patient state [2]**

| Patients | Fusion[*]— % (no. of instances) | GPS[†]— % (no. of instances) | Accelerometer[‡]— % (no. of instances) |
|---|---|---|---|
| p0101 | 70 (70) | 77 (26) | 75 (70) |
| p0102 | 84 (46) | 82 (34) | 76 (46) |
| p0201 | 68 (38) | 77 (36) | 68 (38) |
| p0302 | 82 (60) | 92 (47) | 66 (60) |
| p0502 | 71 (58) | 85 (28) | 72 (58) |
| p0602 | 77 (31) | 71 (31) | 66 (21) |
| p0702 | 74 (42) | 77 (31) | 73 (42) |
| P0802 | 79 (62) | 89 (37) | 77 (62) |
| p0902 | 83 (35) | 85 (35) | 70 (35) |
| p1002 | 68 (43) | 79 (22) | 71 (43) |
| **Mean** | **76** | **81** | **72** |

[*] Fusion: 70.3% recall and 74% precision

[†] Location: 81.7% recall and 80.8% precision

[‡] Accelerometer: 62.9% and 64.8% precision

**Detecting patients' change of state.** We also investigated detection of a state change without explicit recognition of the new bipolar state. This was important, because detecting patients' change of state (which can indicate onset of an episode) can lead to a visit to the clinic and allow early intervention.

In this approach, we built a model of a single patient state. All points falling outside this model were classified as a change. The approach of starting with a single default state has an advantage in that a new patient who comes to the clinic can be given a device to measure state changes as soon as initial data has been collected for their current state.

After fitting a multivariate Gaussian distribution to the default state and establishing distance measures using Mahalanobis distance, we evaluated the model. The results showed that for each patient, state changes could be detected with an average precision of 96% and average recall of 94% (see Table 2).

**Table 2. State change detection [2]**

| Patients | Recall (%) | Precision (%) |
|---|---|---|
| p0101 | 91.1 | 93.4 |
| p0102 | 86.2 | 96.8 |
| p0201 | 97.3 | 92.9 |
| p0302 | 100 | 93.8 |
| p0502 | 97.8 | 97.6 |
| p0602 | 100 | 87.4 |
| p0702 | 96.8 | 97.1 |
| p0802 | 95.6 | 95.2 |
| p0902 | 100 | 97.1 |
| p1002 | 100 | 91.2 |
| **Average** | **96.5** | **94.2** |

**Analyzing the patient's voice and calls.** It's well known that bipolar episodes cause changes in a patient's voice, such as in prosody and speech fluency [5]. As such, we extended the analysis with data from phone call patterns and sound analysis. We divided the sound analysis into speech features to understand dyadic communication of the patient with the other person on the line and into voice features to detect emotions [6].

Table 3 shows results of episode detection based on analysis of phone calls, sound analysis, and the fusion of both. (Note that phone call data was not available for four of the patients.) Speech analysis and phone call data didn't perform as well in episode prediction compared to the location (GPS) data shown in Table 1. We also investigated how well the fusion of all available data could detect a patient's state change.

**Table 3. Accuracy, recall, and precision for phone, sound, and sensor-fusion [6]**

| | Phone | | Sound | | Fusion | |
|---|---|---|---|---|---|---|
| Patients | Percentage correct (no. of total instances) | Recall/ precision | Percentage correct (no. of total instances) | Recall/ precision | Percentage correct (no. of total instances) | Recall/ precision |
| p0102 | 75 (46) | 64.4/70.1 | 66 (46) | 51.2/40.4 | 73 (46) | 58.5/680 |
| p0201 | 62 (38) | 52.5/53.2 | 68 (32) | 60.8/62.0 | 71 (38) | 58.7/66.4 |
| p0302 | 71 (60) | 62.0/63.6 | 74 (60) | 64.5/52.0 | 71 (60) | 60.7/65.3 |
| p0602 | 36 (35) | 33.9/35.0 | 76 (35) | 68.5/78.7 | 65 (35) | 57.0/48.0 |
| p0902 | 68 (41) | 63.7/65.3 | 71 (41) | 68.5/68.5 | 68 (41) | 62.3/65.5 |
| p1002 | 65 (37) | 78.7/69.4 | 65 (37) | 54.0/40.8 | 65 (40) | 53.3/41.4 |
| **Average** | **66** | **61/58** | **70** | **60/59** | **69** | **52/55** |

**Detecting state change by fusing sensor modalities.** We tested a set of fusion strategies—namely, logical AND, OR, and our own *weighted fusion* [6]. The results shown in Table 4 reveal that an AND concatenation (meaning a state change is detected only when all sensor modalities detect change) wasn't very precise. Furthermore, it didn't detect many changes (low recall). This is to be expected, considering that features come from four distinct sensors; the AND strategy would imply that behavior changes must be significantly reflected in all the sensor modalities.

Using an OR concatenation, almost all changes were detected (very high recall); however, there were a number of false alarms (lower precision). By applying the self-designed weighted-fusion concatenation, both recall and precision were very high, meaning that almost all changes were detected with almost no false alarms.

**Table 4. State change detection using fused modalities [6]**

|  | Recall (%) | Precision (%) |
|---|---|---|
| A+L[*] weighted fusion | 96.4 | 94.5 |
| All-in AND fusion | 42.87 | 61.18 |
| All-in OR fusion | 92.15 | 70.28 |
| All-in weighted fusion | 97.36 | 97.19 |

\* Accelerometer and location data

Using all features improved state change detection with respect to the previously described accelerometer and location data. Table 4 shows that the best accuracy can be achieved in fusing all sensor modalities, meaning that our analysis considered all disease-relevant aspects of behavior.

## Conclusion

The work described here is a first step toward using smartphone data to monitor symptoms of mental disorders. One of the important aspects of this work is the possibility of the early detection of changes in a patient's state with high precision and recall, facilitating timely intervention and thus leading to better treatment outcomes. The fact that this study was carried out using a large, real-world dataset, recorded during the daily lives of patients, serves as solid evidence of the potential of smartphones in transforming mental health and wellbeing [4].

While these results are highly encouraging, we plan to investigate whether similar results can be obtained with a higher number of patients, monitored over a longer period of time. We've already begun the initial steps in this direction through the Nympha-MD (www.nympha-md-project.eu), a follow-up European project.

## Inset

Smartphone sensors are being repurposed to measure phenomena beyond their original designation. For example, accelerometer sensors (originally designed for adapting the user interface based on device orientation) are increasingly being used to monitor activities (such as in Google Fit and Apple HealthKit). Furthermore, our work at CREATE-NET has shown that accelerometer sensors can be even used to detect stress [4]. The trend of using smartphone sensors for behavior monitoring will continue, as more smartphones are equipped with increasing sensing modalities. This will enable objective measurement of a multitude of aspects of human behavior, thus lessening reliance on individuals' memory to recall their past behavior—a current practice in psychiatric assessment based on self-reports.

## Acknowledgments

*This work was carried out in the context of the MONARCA project (www.monarca-project.eu).*